\documentclass[twocolumn,secnumarabic,amssymb,nobibnotes,showpacs,aps,prd]{revtex4-1}

\usepackage{hyperref}
\usepackage{graphicx}

\begin{document}

\title{\Large{Leptonic and Digamma decay Properties of S-wave quarkonia states }}

\author{\large Manan Shah}
\email[]{mnshah09@gmail.com}

\author{\large Arpit Parmar}
\email[]{arpitspu@yahoo.co.in}
\author{\large P C Vinodkumar}
\email{pothodivinod@yahoo.com}
\affiliation{Department of Physics, Sardar Patel University,Vallabh Vidyanagar, INDIA.}
\date{\today}%

\begin{abstract}
Based on Martin like potential, the S-wave masses of quarkonia have been reviewed. Resultant wave functions at zero inter quark separation are employed to compute the hyperfine splitting of the nS states and the leptonic and digamma decay widths of $n{^3S_1}$ and $n{^1S_0}$ states of quarkonia respectively. Analysis on the level differences of S-wave excited states of quantum mechanical bound systems show a systematic behaviour as n-increases. In view of such systematic behaviour expected for quarkonia, we observe that   $Y(4263)$ and $X(4630)$ $1^{--}$ states are closer to the 4S and 6S states while $\psi(4415)$ and $Z(4430)$  are closer to the 5S state of $c\bar{c}$ systems. Similarly we find $\Upsilon (10865)$ is not fit to be the 5S state of $b\bar{b}$ system. while $Y_b (10880)$ observed by Belle or (10996) observed by Babar fit to be the 6S state of bottonia. Our predicted leptonic width, 0.242 keV of $\Upsilon (10579, 4S)$ is in good agreement with the experimental value of 0.272 $\pm$ 0.029 keV. We predict the leptonic widths of the pure 5S and 6S states of upsilon states as 0.191 keV and 0.157 keV respectively. In the case of charmonia, we predict the leptonic widths of the 4S, 5S and 6S states as 0.654 keV, 0.489 keV and 0.387 keV respectively.
\end{abstract}
\maketitle

\subsection*{Introduction}
The recent experimental observations particularly in quarkonia sector have generated renewed interest in the study of hadron spectroscopy \cite{CLEO1,CLEO,PDG2010,babar}. The discovery of the $\eta_b \ (1S)$ state \cite{babar,CLEO1} (BaBar and CLEO collaboration) and $\eta_c \ (2S)$ state and many high precision experimental observations of various hadronic states \cite{PDG2010} have necessitated reconsideration of the parameters involved in the previous studies \cite{Meinel,Gray}. The spectroscopic parameters like the interquark potential parameters that provide the masses of the bound states and the corresponding wave functions obtained from the phenomenology are detrimental in the predictions of their decay widths. Most of the existing theoretical values for the decay rates are based on potential model calculations that employ different types of interquark potentials \cite{D.Ebert,Horace,Guo,N.Brambilla}.\\

Till recently, all that was known above the $D\bar{D}$ threshold was the four vector states $\psi(3770)$, $\psi(4040)$, $\psi(4160)$, $\psi(4415)$. The new renaissance in hadron spectroscopy has come from the recent discovery of the large numbers of new states  X, Y, Z \cite{N.Brambilla-1,Chen1,Chen,Kai,Q.He,T.E.}. The challenges paused  by these new states include the right identification with the proper $J^{PC}$ values and their decay modes.\\

Eventhough the spectroscopy of quarkonium states are well recorded experimentally, the S-wave masses of charmonium states beyond 3S and the bottonium states beyond 4S are still not very well resolved. There seemed to be mixing of other resonances nereby. For example The $1^{--}$ states such as $\psi (3770)$, $Y(4008)$, $Y(4260)$, $Y(4360)$, $X(4630)$, $Y(4660)$, $\Upsilon(10865)$, $\Upsilon(11020)$, $Y_b (10880)$ etc,  may be the quarkonia states either with or without mixing with the nearby resonance states. For instance, $\Upsilon (11020)$ state has recently been analysed to be a mixed bottonium $\Upsilon (6S)$ and $\Upsilon (5D)$ states with mixing angle of $\theta = 40^o \pm 5^o$ \cite{Badalian}.  \\

\paragraph* {}
 In this context, we reconsider the $\Upsilon(nS)$  states of bottonium and $\psi(nS)$  states of charmonium to study their properties. The spectroscopic parameters deduced using a phenomenological approach  will be employed to compute the decay properties such as the leptonic and di-gamma decay widths  with no additional parameters.\\

\subsection*{Methodology}
It has been shown that a purely phenomenological approach to the nonrelativistic potential-model study of $\Upsilon$ spectra and $\psi$ spectra can lead to a static non-Coulombic Power-law potential of the form \cite{a,b}
\begin{equation}\label{eq:1.1}
V(r)=\lambda r^\nu+V_0
\end{equation}
where $\nu$ is close to 0.1 and $\lambda > 0$.

Following general quantum mechanical rules as discussed in \cite{quigg},
the binding energy of a system with reduced mass $\mu$ in a power law potential, $ \lambda r^\nu $ is given by
\begin{eqnarray}\label {eq:1.3}
E_{nl}=\lambda^{2/(2+\nu)}&&\left(2\mu\right)^{-\nu/(2+\nu)}\nonumber\\&&\left[A(\nu) \left(n+\frac{l}{2}-\frac{1}{4}\right)\right]^{2\nu/(2+\nu)}
\end{eqnarray}
and the corresponding square of the probability amplitude of the S-waves at the zero separation of the quark-antiquark system is given by
\begin{eqnarray}\label {eq:1.5}
|\psi_{n}(0)|^2&=&\frac{1}{2\pi^2}\left(\frac{2\mu\lambda}{\hbar^2}\right)^{3/(2+\nu)} \frac{\nu}{(2+\nu)}\nonumber \\&&[A(\nu)]^{3\nu/(2+\nu)}\left(n-\frac{1}{4}\right)^{2(\nu-1)/(2+\nu)}
\end{eqnarray}
where
\begin{equation}\label {eq:1.4}
A(\nu)=\left[2\nu\sqrt{\pi}\ \Gamma\left(\frac{3}{2}+\frac{1}{\nu}\right)\right]/\Gamma(1/\nu),\ \ \
\ \ \nu>0.
\end{equation}
The nonrelativistic Schrodinger bound-state mass (spin average mass) of the $Q\bar{Q}$ ($Q\in b,c$) system follows as
\begin{equation}\label {eq:1.6}
M_{SA}=2m_Q+V_0+E_{nl}
\end{equation}
For the hyperfine split we have considered the standard one gluon exchange interaction \cite{Rai}.
 Accordingly, the hyperfine mass split for the S-wave is given by
\begin{equation}
 \Delta M = A_{hyp} |\psi_{n}(0)|^2/m_Q^2.
\end{equation}
 The b and c quark mass parameters $m_b$ and $m_c$ are taken as 4.67 GeV and 1.27 GeV respectively as given in PDG \cite{PDG2010}.
The vector $\Upsilon(nS)$, $\psi(nS)$ and the pseudoscalar $\eta_b(nS)$, $\eta_c(nS)$ masses are obtained by adding $\Delta M/4$ and $-3\Delta M/4$ respectively to the corresponding spin average mass of the nS state given by eq.(\ref {eq:1.6}).
A fit to this mass formula using the experimental masses of $\Upsilon(1S,2S)$ and the newly discovered $\eta_b(1S)$ states provides us the potential parameters $\lambda , V_0$ and the hyperfine parameter  ($A_{hyp}$) in the case of bottonium system. Similarly, a fit to this mass formula using the experimental masses of $\psi(1S,2S)$ and  $\eta_c(1S)$ states provides us the potential parameters $\lambda , V_0$ and the hyperfine parameter  ($A_{hyp}$) of the charmonium systems. The predicted $\Upsilon (nS)$ , $\eta_b(nS)$ and $\psi (nS)$ , $\eta_c(nS)$ states for n $\geq$ 2 are presented in Table \ref{tab11} and \ref{tab22} respectively.\\

\begin{figure}[h]
\includegraphics[height=3in,width=3.5in]{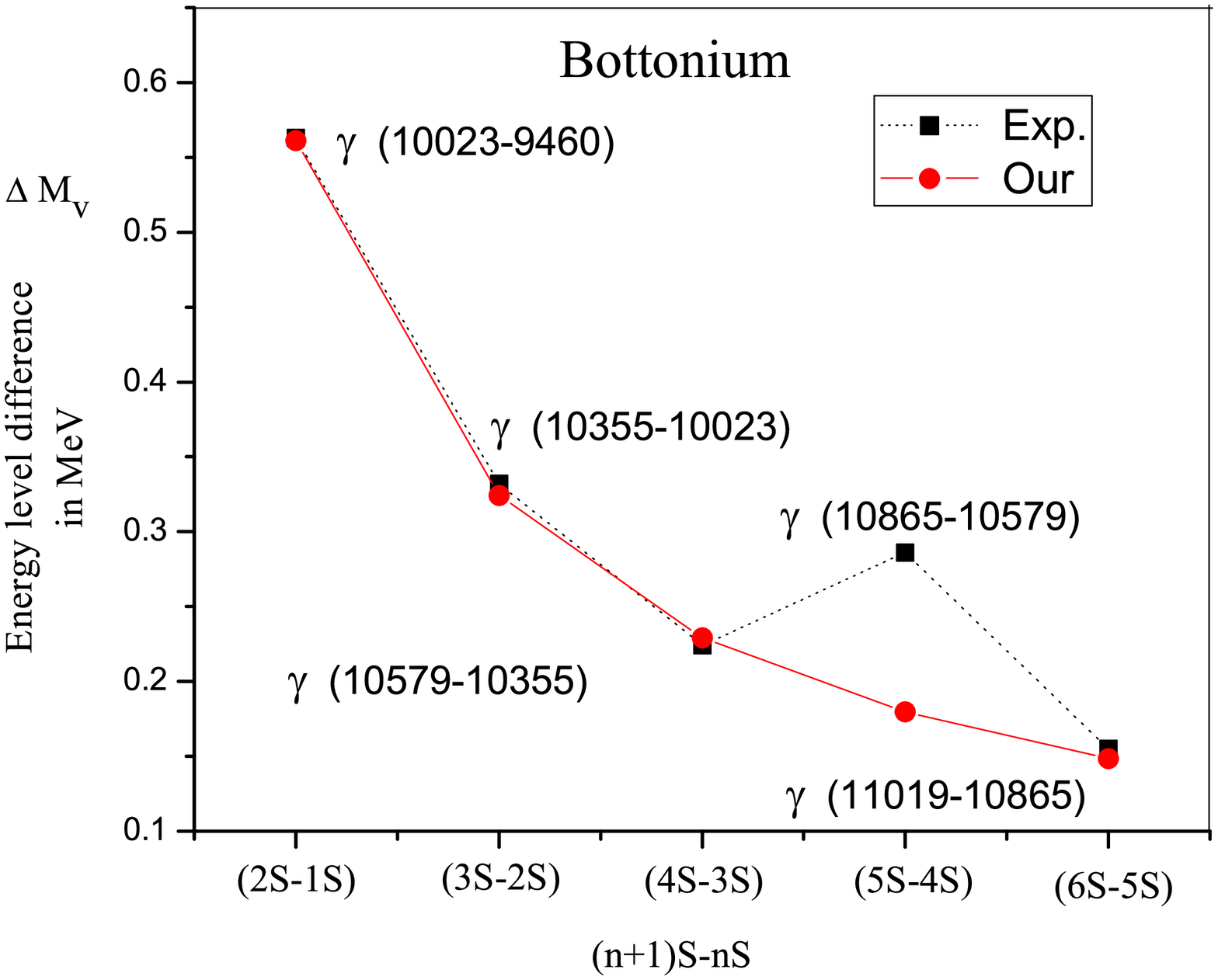}
 \caption{Behavior of energy level shift of the (n+1)S$-$nS states}\label{fig:1}
\end{figure}

\begin{figure}[h]
\includegraphics[height=3in,width=3.5in]{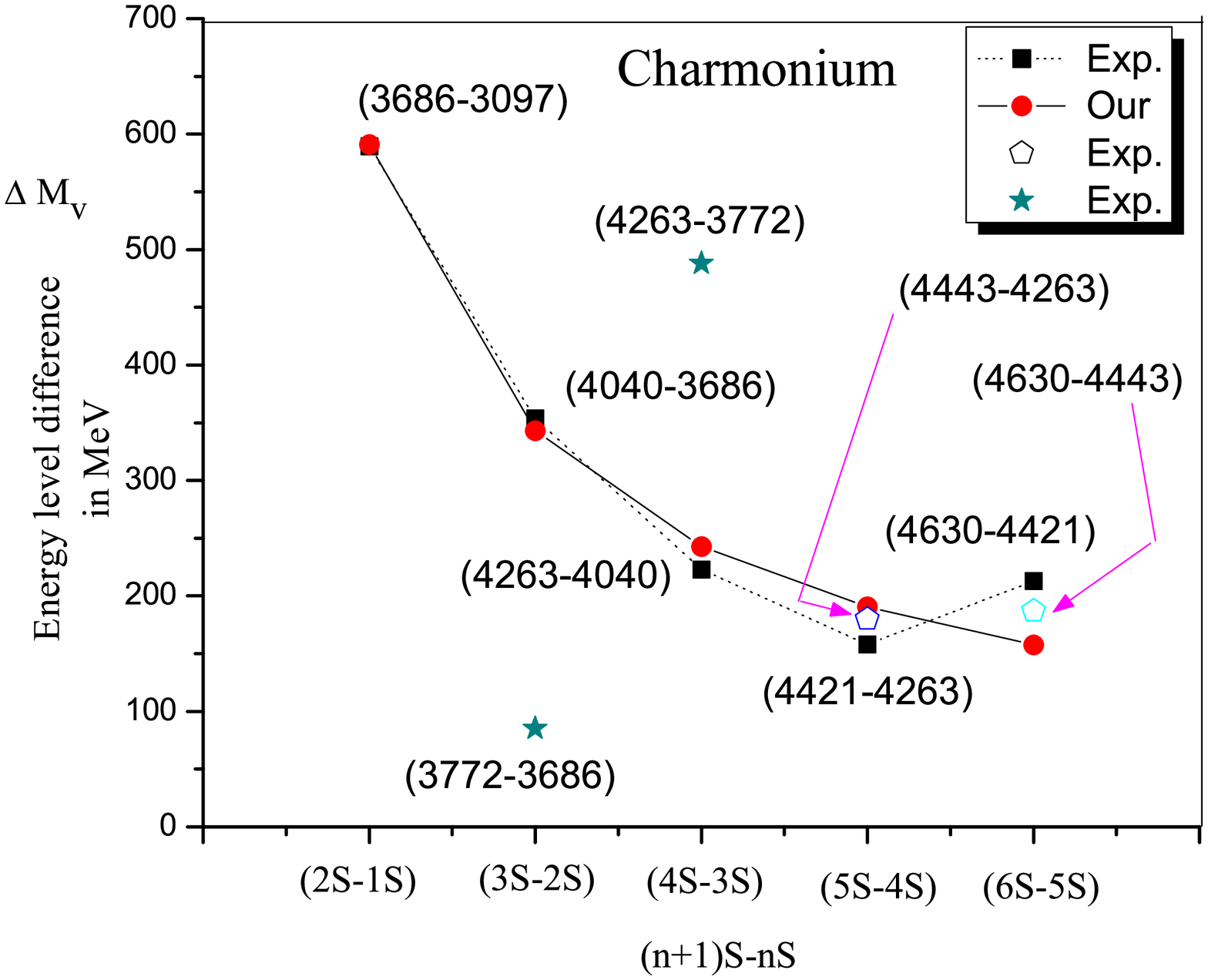}
\caption{Behavior of energy level shift of the (n+1)S$-$nS states}\label{fig:2}
\end{figure}

To compare and identify our predicted states with the respective experimental masses, the PDG  \cite{PDG2010} average values as well as some of the  $1^{--}$ states of X,Y,Z \cite{N.Brambilla-1} are considered for the energy level difference,  $\Delta M = \{\Gamma(n+1)S- \Gamma (nS)\}$. These values are plotted against the (n+1)S-nS for n$=$ 1 to 5 in the case of bottonia and charmonia in FIG.(\ref{fig:1}) and (\ref{fig:2}) respectively. It is expected that the excited states must follow a specific trend line representing its characteristic spectral property. So we compare our predicted states with those which are closer to the systematic expected behaviour shown by the solid line. The states which are widely off from the expected behaviour are then identified as either mixed (disturbed) states or exotic states.


\begin{table}[b]
\begin{center}
\caption{Results for $b\bar{b}$ spectrum }\label{tab11}
\begin{tabular}{ccccccc}
\hline
$nS$ & $M_V$   & $M_V$ & $M_V$ & $M_P$  & $M_P$ & $M_P$\\
    &   $(MeV)$ & $(MeV)$ & $(MeV)$ &  $(MeV)$ & $(MeV)$ & $(MeV)$\\
    &    [our]     &  \cite{1}& Exp.& [our] & \cite{1} & Exp.\cite{PDG2010}\\
\hline\hline
$1S$&	9460.43   &  9460.38  & $\Upsilon$(9460) \cite{PDG2010}   &9392.43    & 9392.91  & $\eta_b$(9391)\\
$2S$&	10021.60  & 10023.3  &  $\Upsilon$(10023) \cite{PDG2010}   &9989.15    & 9987.42  &  -\\
$3S$&	10345.72  & 10364.2  &  $\Upsilon$(10355) \cite{PDG2010}    &10323.46   & 10333.9  &  -\\
$4S$&	10574.91  & 10636.4  &  $\Upsilon$(10579) \cite{PDG2010}    &10557.86   & 10609.4  &  -\\
$5S$&	10754.74  &   -      &  $\Upsilon$(10860) \cite{PDG2010}    &10740.81   & -        &  -\\
$6S$&	10903.15  &    -     &  $\Upsilon$(11020) \cite{PDG2010}    &10891.33   & -        &  -\\
    &             &    \ \ \ \ \ \ {$^{OR}$}&  $Y_b$(10888) \cite{Chen1,Chen} &\\
    &             &    \ \ \ \ \ \ {$^{OR}$}&  $\Upsilon$(10996) \cite{B.Aubert} &\\
\hline
\end{tabular}
\cite{PDG2010}$\rightarrow$ PDG (2010); \cite{Chen1,Chen}$\rightarrow$ Belle (2010); \\ \cite{1}$\rightarrow$ Radford \& Repko (2011); \cite{B.Aubert} $\rightarrow$ Babar (2009).
\end{center}
\end{table}

\begin{table}[h]
\begin{center}
\caption{Results for $c\bar{c}$ spectrum }\label{tab22}
\begin{tabular}{ccccccc}
\hline
$nS$ & $M_V$   & $M_V$ & $M_V$ & $M_P$  & $M_P$ & $M_P$\\
    &   $(MeV)$ & $(MeV)$ & $(MeV)$ &  $(MeV)$ & $(MeV)$ & $(MeV)$\\
    &    [our]     &  \cite{2007}& Exp. & [our] & \cite{2007} & Exp.\cite{PDG2010}\\
\hline\hline
$1S$&	3097.14  & 3096.92 & $J/\psi$(3097) \cite{PDG2010}  &2980.47   & 2981.7  & $\eta_c$(2980)\\
$2S$&	3687.91  & 3686.1  & $\psi$(3687) \cite{PDG2010}    &3631.97   & 3619.2  & $\eta_c$(3637)\\
$3S$&	4030.75  & 4102.0  & $\psi$(4040) \cite{PDG2010}    &3992.39   & 4052.5  &  -\\
$4S$&	4273.51  & 4446.8  & $Y$(4260) \cite{Yuan,Aubert1}  &4244.13   & -       &  -\\
$5S$&	4464.14  &    -    & $\psi$(4415) \cite{PDG2010}    &4440.15   & -        &  -\\
    &	         & \ \ \ \ \ \ {$^{OR}$}& $Z$(4430) \cite{Mizuk,Choi}    &      & &   \\
$6S$&	4621.55  &    -    & $X$(4630) \cite{BES08}         &4601.18   & -        &  -\\
\hline

\end{tabular}
\cite{PDG2010}$\rightarrow$ PDG (2010); \cite{2007} $\rightarrow$ Radford \& Repko (2007);\\
\cite{Yuan} $\rightarrow$ Belle (2007); \cite{Aubert1} $\rightarrow$ Babar (2005);\\
\cite{BES08} $\rightarrow$ BES (2008); \cite{Mizuk} $\rightarrow$ Belle (2009);\cite{Choi} $\rightarrow$ Belle (2008).


\end{center}
\end{table}
\begin{table}[b]
\begin{center}
\caption{The leptonic widths of the $\Upsilon (nS)$ and the di-gamma widths of $\eta_b(nS)$ states}\label{tab3}
\begin{tabular}{cccccccc}
\hline
$(nS) $ &  $\Gamma^{l^+l^-} $ &  $\Gamma^{l^+l^-} $ & $\Gamma^{l^+l^-}$ & $\Gamma^{\gamma\gamma} $ & $\Gamma^{\gamma\gamma} $ & $\Gamma^{\gamma\gamma} $\\
& (keV)&(keV)&(keV)&(keV)&(keV)&(keV)\\
& [our] & \cite{1} &{Exp.}& [our]& \cite{Schuler}& \cite{Mohammad}\\
\hline\hline
$1S$ 	 &	1.207	& 1.33& 1.34$\pm$0.018 \cite{PDG2010} &	0.497 & 0.460  &0.580 \\
$2S$ 	 &	0.513	& 0.62& 0.612$\pm$0.011 \cite{PDG2010} &	0.209 &	0.20  &-\\
$3S$ 	 &	0.330   & 0.48& 0.443$\pm$0.008 \cite{PDG2010} &	0.134 &	-&-\\
$4S$ 	 &	0.242	& 0.40& 0.272$\pm$0.029 \cite{PDG2010} &	0.098 &	-&-\\
$5S$     &	0.191	&   - & 0.310$\pm$0.07 \cite{PDG2010}  &	0.077 &	-&-\\
$6S$     &	0.157	&   - & 0.130$\pm$0.030 \cite{PDG2010} &	0.064 &	-&-\\
\hline
\end{tabular}
\end{center}
\end{table}

\subsection*{Leptonic and di-gamma decay widths of Bottonium and charmonium states}

Apart from the masses of the lowlying states, the hyperfine splits due to chromomagnetic interaction and the right behaviour of the wave
function that provides as the correct predictions of the decay rates are important features of any
successful model. Accordingly,the radial wave functions of the identified nS states of quarkonia ($c\bar{c}, b\bar{b}$) obtained from eqn.(\ref{eq:1.5}) are employed to predict the leptonic and digamma widths of the vector $1^{--}$ and $0^{-+}$ states respectively. The leptonic decay widths with the radiative correction of $\Upsilon (nS)\rightarrow {l^+l^-}$ and $\psi (nS)\rightarrow {l^+l^-}$ are computed as \cite{Ajay2008,Ajay2005,Kwong1988}
\begin{equation}\label{eq:vwgamma2}
\Gamma^{l^+l^-}=\frac {16 \pi \alpha_e^2 e_Q^2}{M_V^2}|\psi^2_{nl}(0)| \left[1 -\frac{16}{3 \pi}
\alpha_s\right]
\end{equation}
And the di-gamma (two photon) decay widths of $\eta_b(nS)\rightarrow {\gamma\gamma}$ and $\eta_c(nS)\rightarrow {\gamma\gamma}$  with radiative correction are obtained as

\begin{equation}
\Gamma^{\gamma\gamma}= \frac{48 \pi \alpha_e^2 e_Q^4}{M_P^2} \ |\psi^2_n(0)| \left[1-\frac{\alpha_s}{\pi}\left(\frac{20-\pi^2}{3}\right)\right]
\end{equation}
$\alpha_e $ is the electromagnetic coupling constant and $\alpha_s $ is the strong coupling constant. The predicted results in the case of bottonia and charmonia are tabulated in Table \ref{tab3} and \ref{tab4} respectively.\\


\begin{table}[h]
\begin{center}
\caption{The leptonic widths of the $\psi (nS)$ and the di-gamma widths of $\eta_c(nS)$ states}\label{tab4}
\begin{tabular}{ccccccc}
\hline
$nS $ &  $\Gamma^{l^+l^-} $ &  $\Gamma^{l^+l^-} $ & $\Gamma^{l^+l^-}$ & $\Gamma^{\gamma\gamma} $ & $\Gamma^{\gamma\gamma} $ & $\Gamma^{\gamma\gamma} $\\
& (keV)&(keV)&(keV)&(keV)&(keV)&(keV)\\
& [our] & \cite{2007} &{Exp.}& [our]& \cite{Schuler}& \cite{Mohammad}\\
\hline\hline
$1S$	 &	4.944		& 1.89&   5.55 $\pm$0.14 \cite{PDG2010} & 10.373 & 7.8    & 11.8 \\
$2S$	 &	1.671		& 1.04&   2.35 $\pm$0.04 \cite{PDG2010} &	3.349 &	3.5    &-\\
$3S$	 &	0.959   	& 0.77&   0.86 $\pm$0.07 \cite{PDG2010} &	1.900 &	-      &-\\
          &             &     &  0.83 $\pm$0.07 \cite{BES08}    &         &        &- \\
$4S$	 &	0.654		& 0.65&          -                      &	1.288 &	-      &-\\
$5S$	 &	0.489		&   - &  0.58 $\pm$ 0.07 \cite{PDG2010} &	0.961 &	-      &-\\
         &             &     &  0.35 $\pm$0.12 \cite{BES08}     &         &   &- \\
$6S$	 &	0.387		&   - &           -      &	0.759 &	-      &-\\
\hline
\end{tabular}
\end{center}
\end{table}



\subsection*{Results and discussion}
We have been able to predict the charmonium and bottonium S-wave masses states which are in good agreement with the reported PDG values as compared to the predicted values of \cite{1}. We have also predicted the $\eta_b(2S-6S)$ states within the mass range 9.989 GeV to 10.891 GeV and $\eta_c(2S-6S)$ states within the mass range 3631.97 GeV to 4601.18 GeV. We hope to find future experimental support in favor of our predictions. With no additional parameters we have been able to predict the leptonic decay widths of $\Upsilon(1S-6S)$ as well as $\psi(1S-6S)$ states which are in good agreement with the known experimental values \cite{PDG2010}.  The predicted di-gamma decay widths of $\eta_b(1S-6S)$ and $\eta_c(1S-6S)$ states would be helpful to identify the $0^{-+}$ resonances in future experiments.\\
 \paragraph* {}
 We do not compare our 5S and 6S states with those of  $\Upsilon$(10860) and  $\Upsilon$(11020) states listed in PDG. They find it difficult to assign it as the 5S and 6S states respectively as there seemed to be mixing of two Briet–Wigner resonances \cite{PDG2010}. The energy level shifts of the (n+1)S$-$nS states clearly indicate that the state  $\Upsilon$ (10860) is far from the expected trend line as shown in FIG. \ref{fig:1}. As expected it is seen from table \ref{tab3} and \ref{tab4}, that the leptonic decay widths falls off as nS goes from 1S to 6S. Such a behaviour is seen even in the reported experimental values except in the case of $\Upsilon (10860)$. Thus we identify $\Upsilon (10860)$ as either a mixed state or exotic, while $Y_b (10888)$ observed very recently by Belle \cite{Chen} and $\Upsilon (10996)$ observed by Babar \cite{B.Aubert} are closer to the predicted 6S states of bottonia.\\

 In the charmonium sector, many vector $(1^{--})$ states have been reported by Belle and Babar \cite{N.Brambilla-1,Chen1,Chen,Kai,Q.He,T.E.}. It is observed that the $1^{--}$ state, $\psi (3770)$ does not fit into the $\psi (3S)$ state, while $\psi (4040)$ belong to the vector 3S state of charmonia. Similarly, the newly discovered vector state $Y (4263)$ \cite{BES08} and $\psi (4415)$ \cite{PDG2010} are closer to the 4S and 5S states of charmonia. Though our predicted 6S state at 4621 MeV is close to the newly discovered $X (4630)$ $1^{--}$ state, the energy level difference as seen from Fig.(\ref{fig:2}) suggests it to be unfit as the 6S state. The behaviour of the energy level difference also suggests $\psi (4415)$ may be a mixed state. We expect $\psi (5S)$ state to be about 200 MeV above $Y (4260)$ state. Such a state with mass of $(4443^{+24}_{-18})$ represented by $Z(4430)$ has been reported by Belle, though its $J^{PC}$ is not yet known \cite{Mizuk,Choi}. Then $X (4630)$ state becomes the right candidate for the 6S state of charmonia. Thus future experimental confirmation of vector charmonia state around 4464 MeV can resolve the issue of the charmonium 5S and 6S vector states. And we predict the leptonic decay widths of $\psi (5S,6S)$ states around 0.49 keV and 0.39 keV as well as that for the $\Upsilon (5S,6S)$ states around 0.19 keV and 0.16 keV respectively. \\



\subsection*{Acknowledgments}

The work is done under UGC Major research project NO. F.40-457/2011(SR).



\begin{thebibliography}{50}
\bibitem{CLEO1}G.Bonvicini,et al.,CLEO collaboration,Phys. Rev.D 81 (2010) 031104.
\bibitem{CLEO} K.M.Ecklund, et al.,CLEO collaboration,Phys. Rev.D 78 (2008) 091501.
\bibitem{babar}B.Auger,et al.,BaBar collaboration,Phys. Rev. Lett. 103(2009)161801.
\bibitem{PDG2010} K Nakamura \emph{et al.} (Particle Data Group), J.Phys.  G, 37,075021 (2010).
\bibitem{Meinel}S.Meinel,RBC Collaboration,UKQCD Collaboration,Phys. Rev. D 79(2009)094501.
\bibitem{Gray}A.Gray,et al.,HPQCD Collaboration,UKQCD Collaboration,Phys. Rev. D 72(2005)094507.
\bibitem{D.Ebert} D.Ebert,R.N.Faustov,V.O.Galkin,Int.Mod. Phys. Lett. A 20(2005) 1887.
\bibitem{Horace} Horace W. Crater, Cheuk-Yin Wong, Peter Van Alstine, Phys. Rev. D 74 (2006) 054028.
\bibitem{Guo} Guo-Li Wang, Phys. Lett. B 653 (2007) 206.
\bibitem{N.Brambilla} N.Brambilla,E.Mereghetti, A. Vairo, Phys. Rev. D 79 (2009)  074002.
\bibitem{N.Brambilla-1} N.Brambilla et al.,Eur.Phys. J. C 71 (2011) 1534.
\bibitem{Chen1}K.F. Chen et al. (Belle Collaboration), Phys. Rev. D 82(2010)091106(R).
               arXiv:0808.2445[hep-ex]
\bibitem{Chen}K.F. Chen et al. (Belle Collaboration), Phys. Rev. Lett. 100(2008)112001.
                arXiv:0710.2577 [hep-ex]
\bibitem{Kai}Kai Yi (for the CDF Collaboration), arXiv:1010.3470 [hep-ex]
\bibitem{Q.He}Q. He et al. (CLEO Collaboration), Phys. Rev. D 74(2006)091104. arXiv:hep-ex/0611021
\bibitem{T.E.}T.E. Coan et al. (CLEO Collaboration), Phys. Rev. Lett. 96(2006)162003 . arXiv:hep-ex/0602034
\bibitem{Badalian}A.M. Badalian, B.L.G. Bakker and I.V. Danilkin. Physics of Atomic Nuclei,73(2010)138.
\bibitem {a} A.Martin, phys. Lett. 93B, 338 (1980).
\bibitem {b} N.Barik and S. N. Jena, Phys. Lett. 97B, 261 (1980).
\bibitem{quigg}C. Quigg and J.L. Rosner, Phys Rep.  56, No. 4 (1979) 167—235.
\bibitem{Rai}Ajay Kumar Rai,Bavin Patel and P C Vinodkumar, Phy.Rev. C78,055202(2008).

\bibitem{Ajay2008} Rai A K, Pandya  J N and  Vinodkumar P C, Eur. Phys. J.  A4, 77 (2008).
\bibitem{Ajay2005}  Rai A K,  Pandya J N  and Vinodkumar P C, J. Phys. G: Nucl. Part. Phys. 31,
                    1453 (2005).
\bibitem{Kwong1988} Kwong W,  Mackenzie P B, Rosenfeld R and Rosener J L, Phys. Rev.  D37, 3210(1988).
\bibitem{1} Stanley F. Radford, Wayne W. Repko,Nucl. phys A 865(2011) 69-75.
\bibitem{B.Aubert}B. Aubert et al. (BABAR Collaboration), Phys. Rev. Lett. 102(2009)012001.
                  arXiv:0809.4120 [hep-ex]
\bibitem{2007} Stanley F. Radford, Wayne W. Repko,Phys. Rev. D 75(2007) 074031.
\bibitem{Schuler} G.A.Schuler,F.A.Berends,R.vanGulik,Nucl. Phy. B 523 (1998)423-438.
\bibitem{Mohammad}Mohammad R. Ahmady and Roberto R. Mendel, Phys. Rev. D 51 (1995)141.

\bibitem{Yuan} C.Z. Yuan et al. (Belle Collaboration), Phys. Rev. Lett. 99(2007)182004.
               arXiv:0707.2541 [hep-ex]
\bibitem{Aubert1}B. Aubert et al. (BABAR Collaboration), Phys. Rev. Lett. 95(2005)142001.
                 arXiv:hep-ex/0506081
\bibitem{BES08} M. Ablikim et al. (BES Collaboration), Phys. Lett. B 660 (2008) 315.
                arXiv:0705.4500 [hep-ex]
\bibitem{Mizuk} R. Mizuk et al. (Belle Collaboration), Phys. Rev. D 80 (2009) 031104.
                arXiv:0905.2869 [hep-ex]
\bibitem{Choi} S.K. Choi et al. (Belle Collaboration), Phys. Rev. Lett. 100 (2008)142001.
               arXiv:0708.1790 [hep-ex]

\end{thebibliography}
\end{document}